\documentclass[12pt]{iopart}
\newcommand{\pT}{\ensuremath{p_{T}}}
\newcommand{\pTtrig}{\ensuremath{p_{T,trig}}}
\newcommand{\pTassoc}{\ensuremath{p_{T,assoc}}}
\newcommand{\deta}{\ensuremath{\Delta\eta}}
\newcommand{\dphi}{\ensuremath{\Delta\phi}}
\usepackage{iopams}
\usepackage{epsfig}
\usepackage{cite}
\begin{document}

\title{Quark Matter 2006: High-\pT{} and jets}

\author{M van Leeuwen}

\address{Lawrence Berkeley National Laboratory, CA 94720, USA}
\ead{mvanleeuwen@lbl.gov}
\begin{abstract}
An overview of new experimental results on high-\pT{} particle
production and jets in heavy ion collisions from the Quark Matter 2006
conference is presented.
\end{abstract}
In these proceedings, I will summarise new results on high-\pT{}
particle production and jet-like correlations from Quark Matter
2006. The large statistic data sample collected at RHIC in p+p, Cu+Cu
and Au+Au provides access to increasingly high \pT{} and allows to
explore the intermediate \pT{} regime in more detail. At the highest
\pT, we explore our understanding of the 'perturbative limit' where
hadron production and energy loss calculations are relatively well
controlled. This allows to extract characteristics of in-medium
fragmentation and ultimately to measure the energy density or even the
temperature of the medium in heavy ion collisions \cite{Liu:2006ug}.
At intermediate \pT, a number of surprising effects has been observed,
such as the large baryon/meson ratio
\cite{star_LamK,Adler:2003cb,star_piprot} and strong modifications of
triggered di-hadron correlations \cite{Adams:2005ph,Adler:2005ee}. An intriguing new result in this
area is the observation of correlated particle production associated
with jets at large rapidity separation $\Delta\eta$. Also of special
interest are jet-like correlations of identified particles, because
these may shed light on the origin of the baryon/meson enhancement at
intermediate \pT.

It should be noted that this field is also developing at SPS: at the
previous Quark Matter and also in this meeting, a number of new
analyses were presented of nuclear modification factors for hadron
production ranging up to $\pT = 4-5$ GeV. New results exist from NA49
\cite{Alt:2005cb}, NA57 \cite{Dainese:2005vk} and
WA98 \cite{reygers_qm06}. So far, it seems that at SPS energies the
Cronin effect plays a large role, also due to the intrinsically
limited \pT-reach at low $\sqrt{s_{NN}}$. 

\section{$R_{AA}$ and the medium density}
\begin{figure}
\epsfig{file=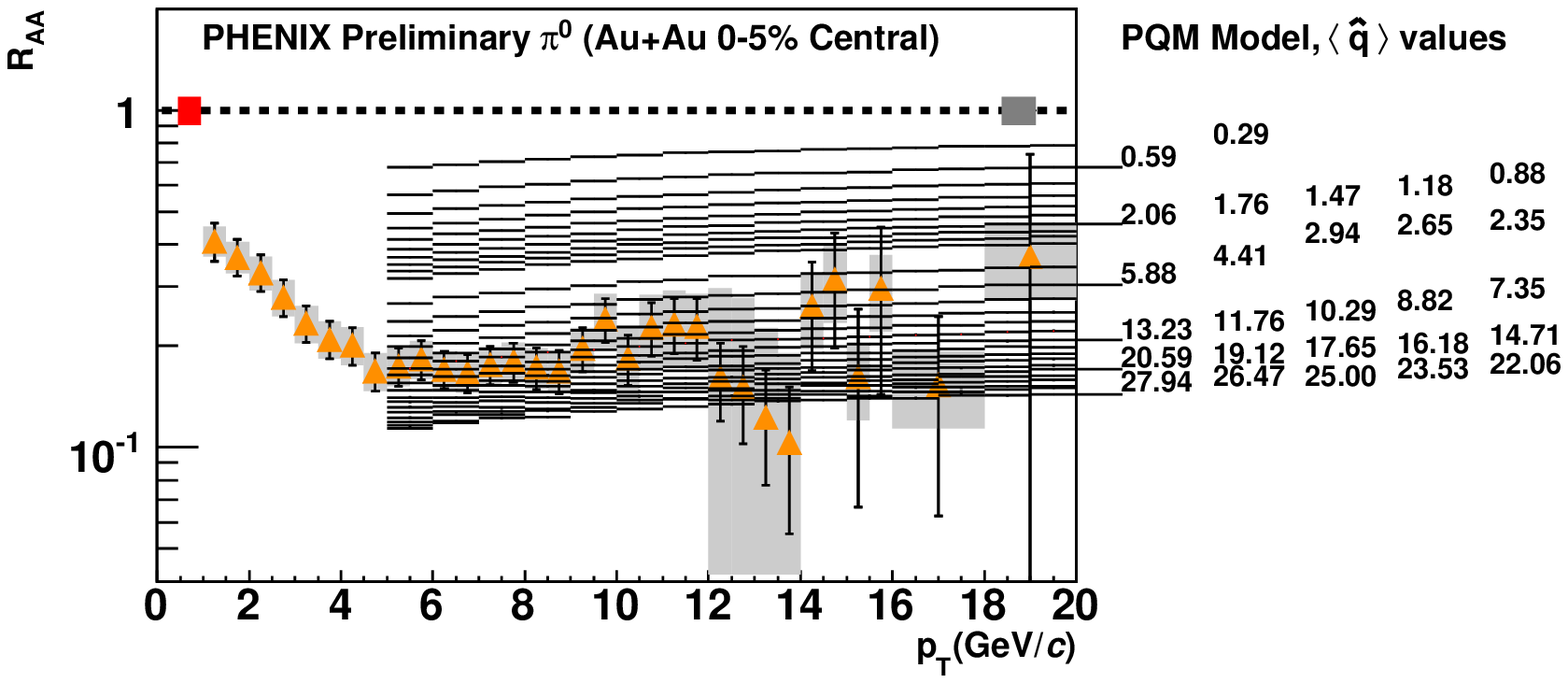,width=0.54\textwidth}%
\epsfig{file=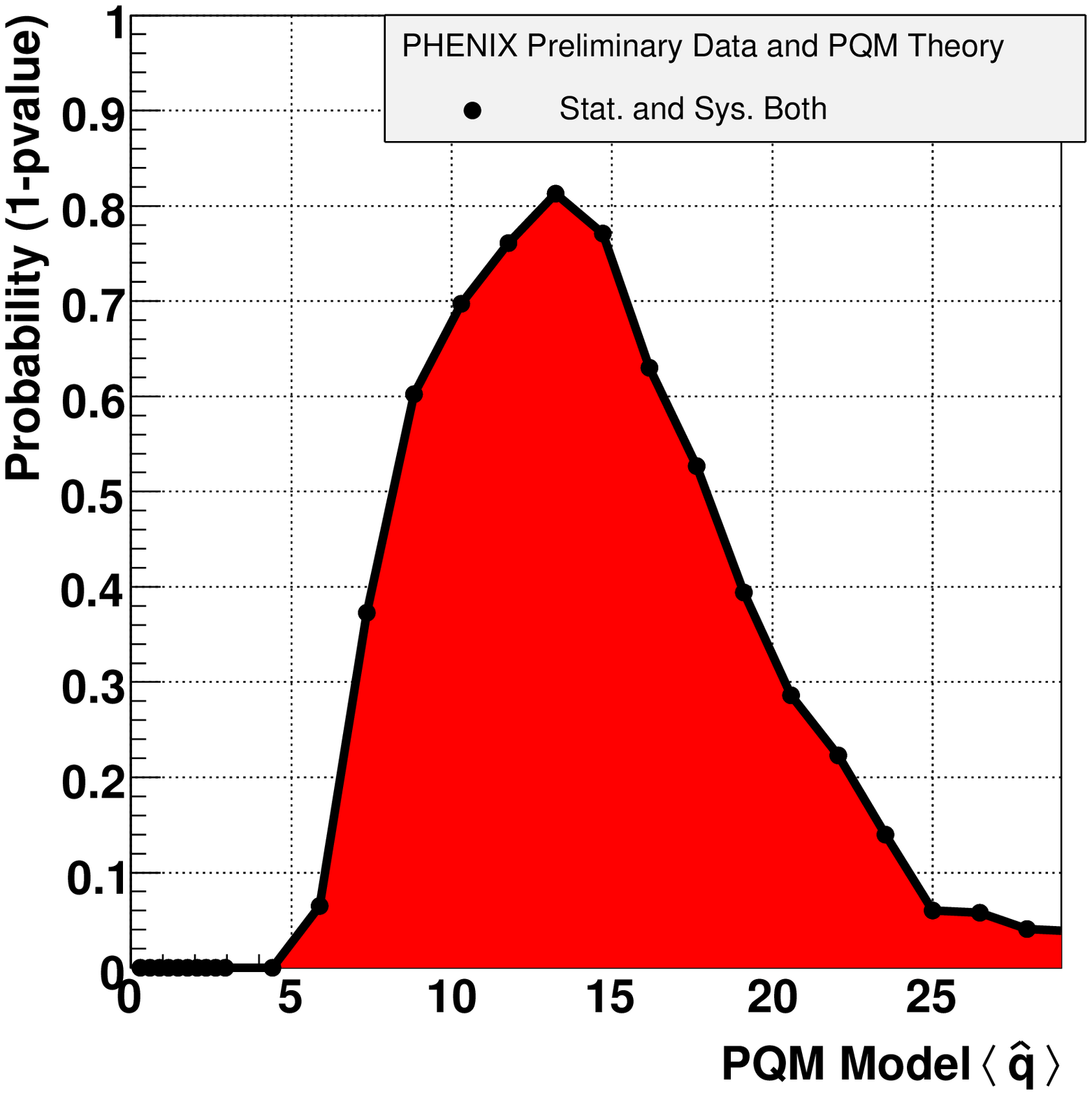,width=0.23\textwidth}%
\epsfig{file=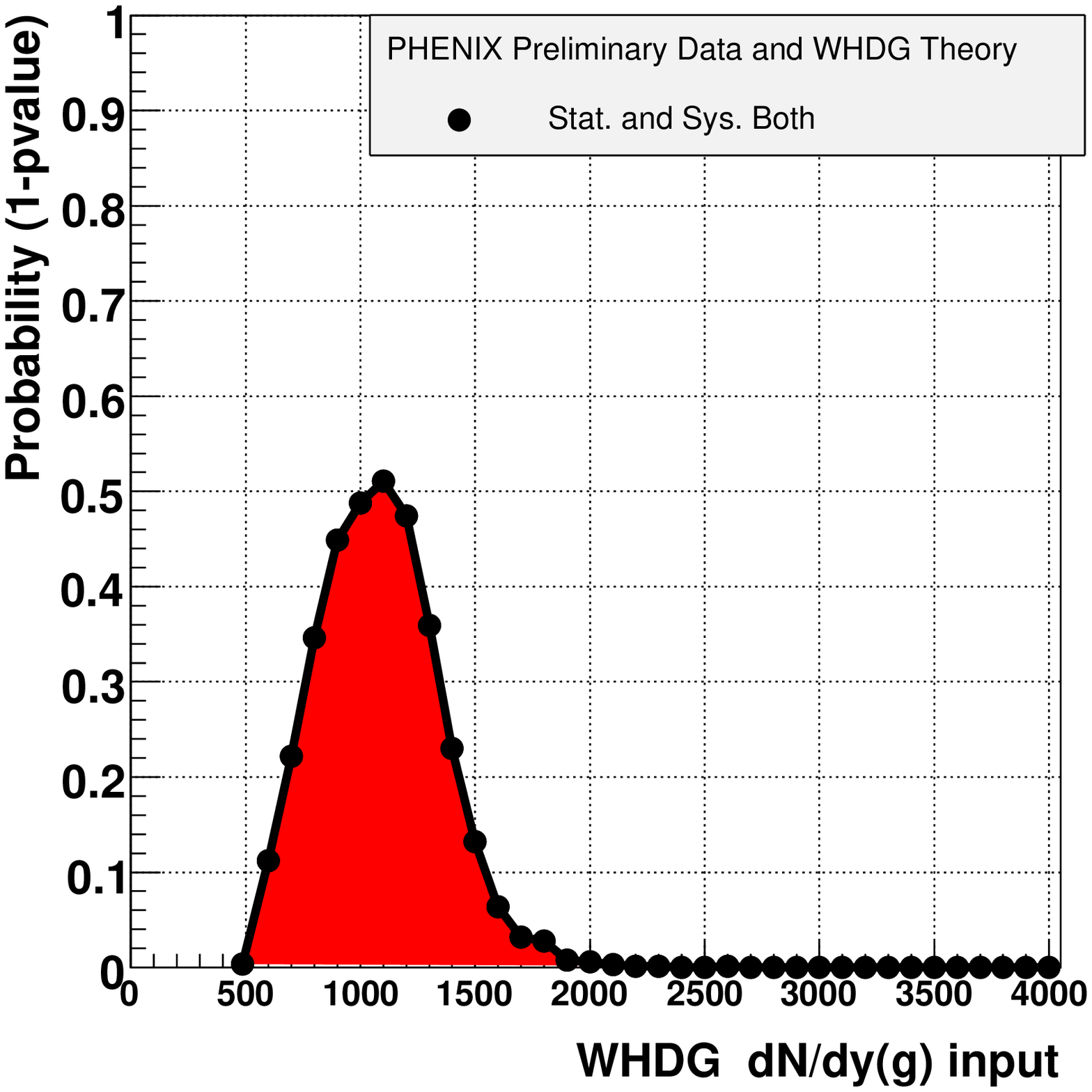,width=0.23\textwidth}
\caption{\label{fig:raa}Left panel: Measurements of the nuclear
  modification factor $R_{AA}$ for $\pi^0$ from PHENIX compared to PQM
  model calculations with different medium density.  Middle and right
  panels: likelihood distribution for data given the medium density,
  using the PQM model (middle) and the GLV
  formalism (right).}
\end{figure}
The suppression of high-\pT{} particle production compared to a
properly scaled p+p reference was the first indication of significant
parton energy loss at RHIC. At this year's conference, a detailed
quantitative confrontation of various models with the measured nuclear
modification factor $R_{AA}$ of $\pi^0$, reaching out to $\pT = 20$
GeV, was presented \cite{lajoie_qm06}. The left panel of
Fig.\ \ref{fig:raa} shows the data and curves from the PQM model
\cite{Dainese:2004te}, using different values of the medium density,
expressed as the time-averaged the transport coefficient
$\hat{q}$. For every value of the medium density, a likelihood value
was calculated using the statistical and systematic uncertainties on
the data. The resulting likelihood distribution is shown in the middle
panel. The right panel shows a similar curve based on the GLV
formalism \cite{Wicks:2005gt}. The extracted limits on the medium
density, for 90\% probability, are $6<\langle \hat{q} \rangle<24$
GeV/fm$^2$ and $600 < dN_g/dy < 1600$ respectively. The model
calculations underlying these predictions are becoming more realistic,
using Woods-Saxon based medium density profiles and fluctuations of
the energy lost by gluon radiation. Such effects may impact the value
of $R_{AA}$, because the energy loss can be large and the kinematic
range at RHIC is limited. It is exciting to see that high-\pT{}
physics at RHIC is entering the quantitative era.

In order to fully constrain the dynamics, more differential
measurements, such as di-hadron correlations, path length dependent
measures (elliptic flow $v_2$ or $R_{AA}$ as a function of the angle
with the reaction plane) and $\gamma$-jet measurements should be
confronted with energy loss calculations. This will provide more insight in
for example the probability distribution for energy loss, which is
different in the multiple-soft scattering (BDMPS) and the
few-hard scatterings (GLV) limits \cite{Salgado:2003gb}, and which is not
well constrained by $R_{AA}$ alone \cite{Renk:2006qg}.

\section{Di-hadron correlations at intermediate and high \pT}

\begin{figure}
\epsfig{file=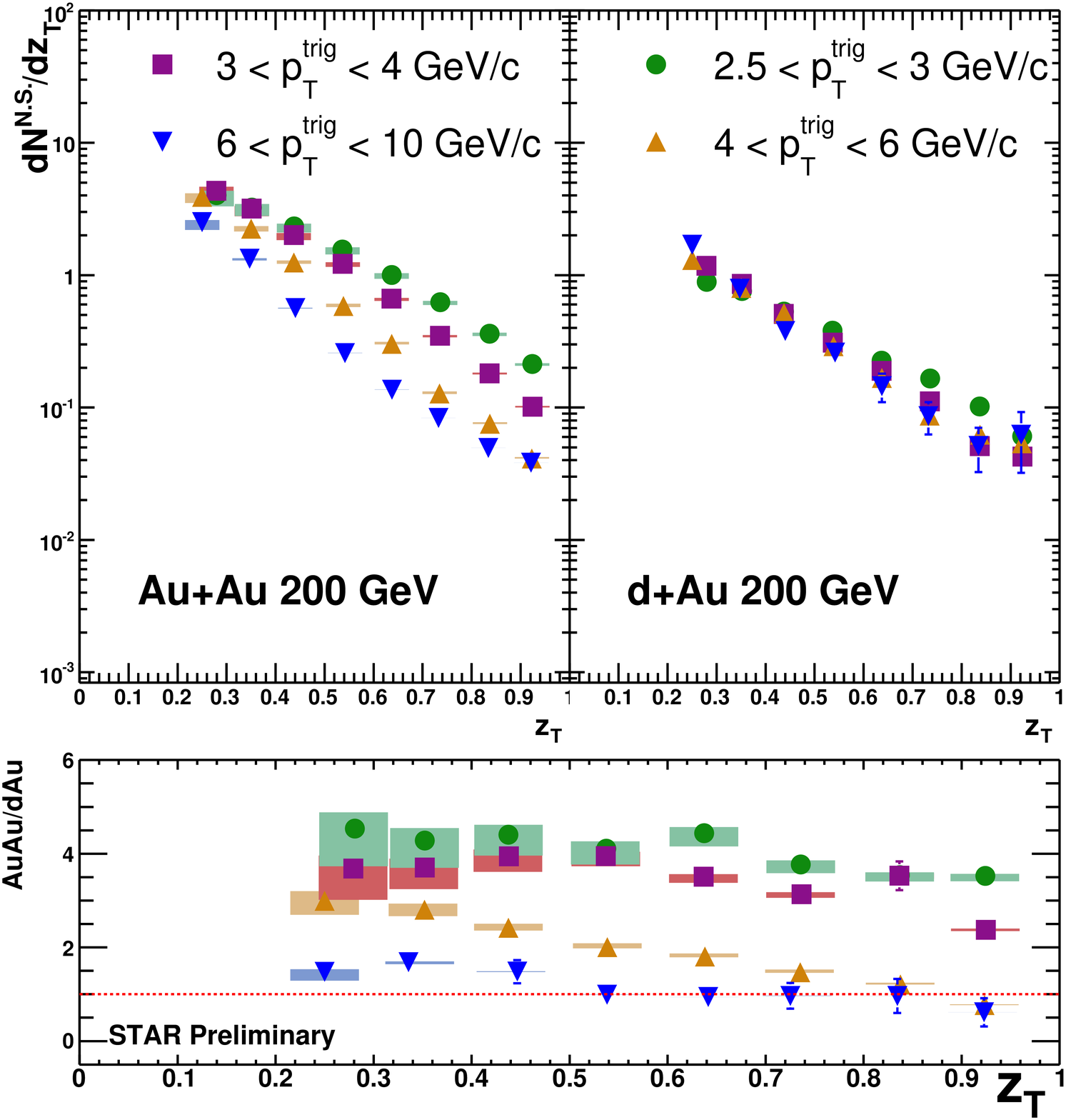,width=0.45\textwidth}\hfill{}
\epsfig{file=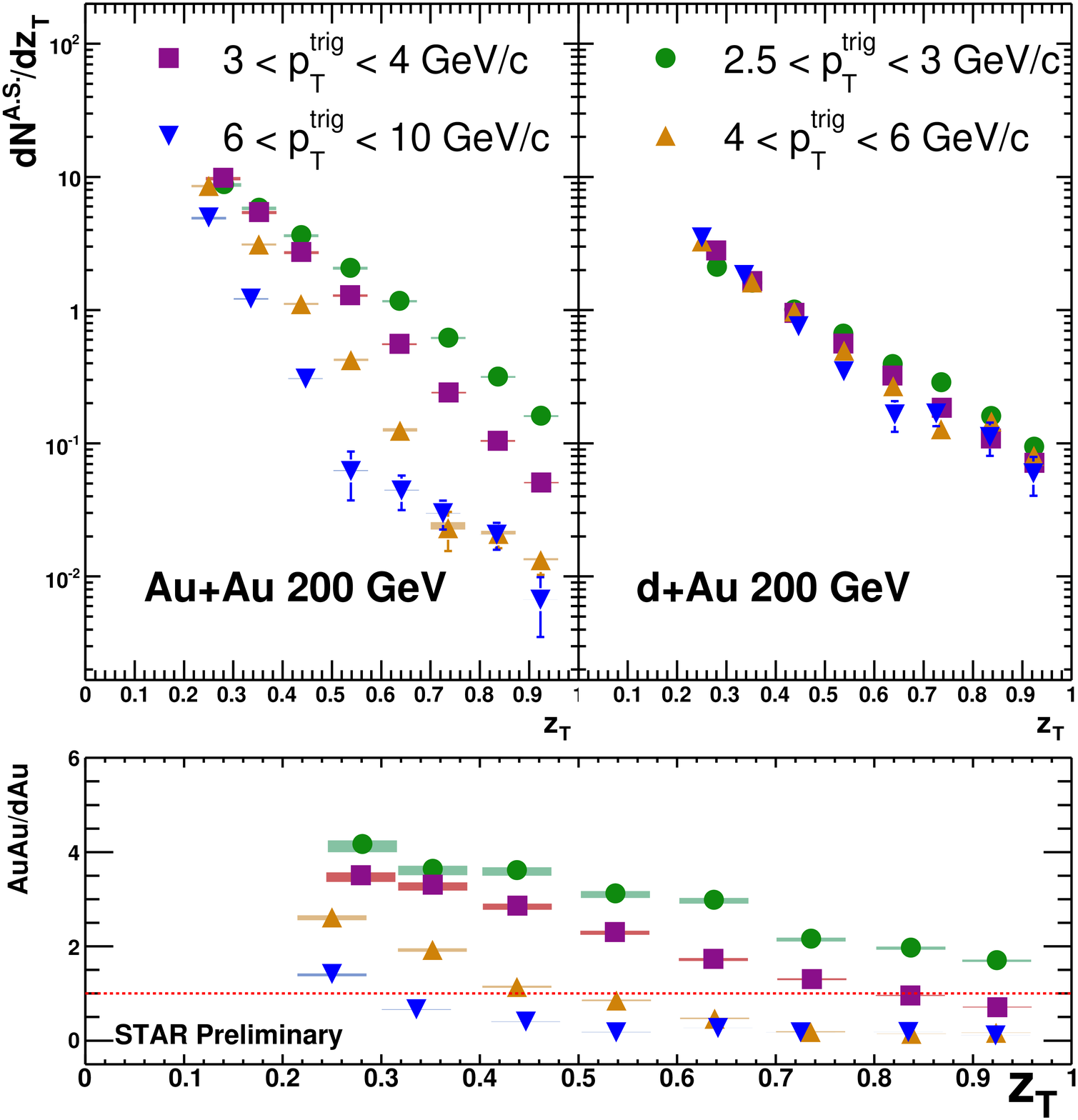,width=0.45\textwidth}
\caption{\label{fig:dihadr_star}Correlated yields on the near side
  ($|\Delta\phi| < 0.9$, left panels) and away side ($|\Delta\phi| >
  0.9$, right panels) as a function of $z_T=\pTassoc/\pTtrig$, for
  several ranges of $p_{T,trig}$. Results from 0-12\% central Au+Au
  and d+Au collisions are shown and the ratio Au+Au/d+Au is presented
  in the lower panels. The shaded bands show the uncertainty due to
  elliptic flow of the background.}
\end{figure}
Fig.\ \ref{fig:dihadr_star} presents an overview of associated yields
with intermediate and high-\pT{} trigger hadrons
\cite{horner_qm06}. The left panel shows the near side yields
($|\Delta\phi| < 0.9$) and the right panel shows the away side yields
($|\Delta\phi| > 0.9$). The results are presented as a function of
$z_T=\pTassoc/\pTtrig$. The associated yields in d+Au collisions show
an approximate scaling in this variable, with the spectra becoming
slightly steeper with increasing \pTtrig. In Au+Au collisions, on the
other hand, a marked increase of the yield with decreasing \pTtrig{}
is observed on both the near and away sides.

On
the near side, the yields in Au+Au collisions are similar to d+Au only
for the highest \pTtrig{} ($6 < \pTtrig < 10$ GeV), with a small
enhancement at the lowest $z_T$. For the lower \pTtrig{},
enhancements up to a factor four are seen in Au+Au collisions. This
enhancement is further discussed in the next section.

On the away side a similar evolution is seen, but with the away-side
yields in Au+Au being suppressed by a factor 4-5 below the d+Au yields
at high \pT. The suppression at high \pTtrig{} is consistent with
expectations from energy loss calculations using medium densities
similar to those used for the calculations that describe $R_{AA}$
\cite{Dainese:2004te,Renk:2006pk}. Due to the counterbalancing biases
from the trigger and associated hadron \pT, di-hadron measurement are
more sensitive to the medium density profile than inclusive
measurements \cite{Renk:2006pk}. At lower \pTtrig, the away-side
yields in Au+Au collisions show an enhancement rather than a
suppression. The evolution from suppression to enhancement is a
function of both \pTtrig{} and \pTassoc{}.

Qualitatively, the observed changes in the correlated yields can be
induced by parton energy loss. For a given trigger hadron \pT{}, a
reduction of the transverse momentum of the leading charged hadron due
to energy loss in Au+Au events could lead to a higher initial jet
energy being selected than in d+Au events. The larger jet-energy would
then lead to an enhanced correlated yield in Au+Au. It is also
possible that the increased yield is due to a jet-induced response of
the medium.

It should also be noted that the strongest modifications of the
correlated yields are seen at the lower \pT{}, where the baryon/meson
ratio is also found to be enhanced compared to expectations from
vacuum fragmentation \cite{star_LamK,Adler:2003cb,star_piprot}. A
natural question therefore is whether the modifications in the
correlated yields and the increased baryon/meson ratio have a common
origin.

\subsection{Near side shape: the ridge}
\begin{figure}
\epsfig{file=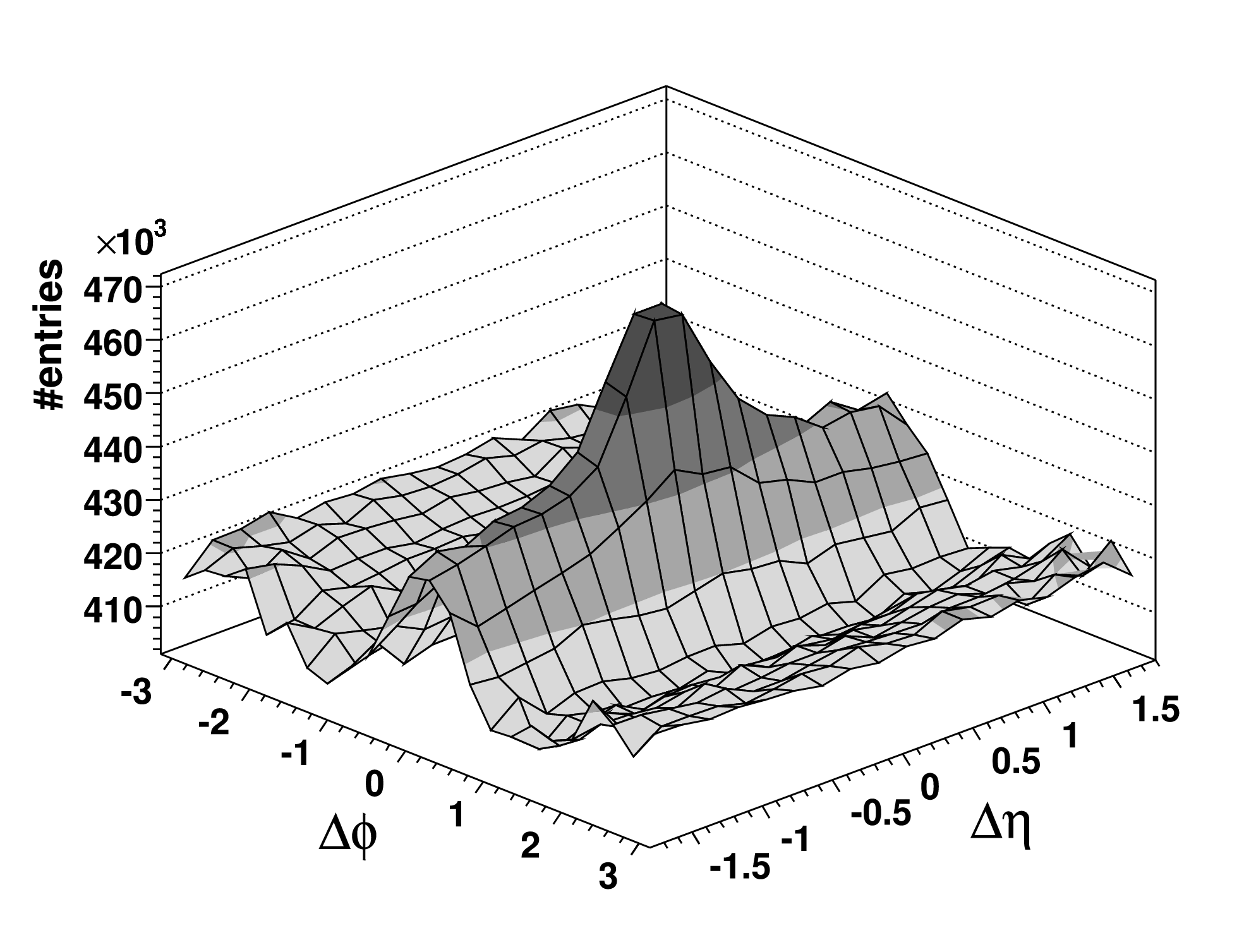,width=0.3\textwidth}%
\epsfig{file=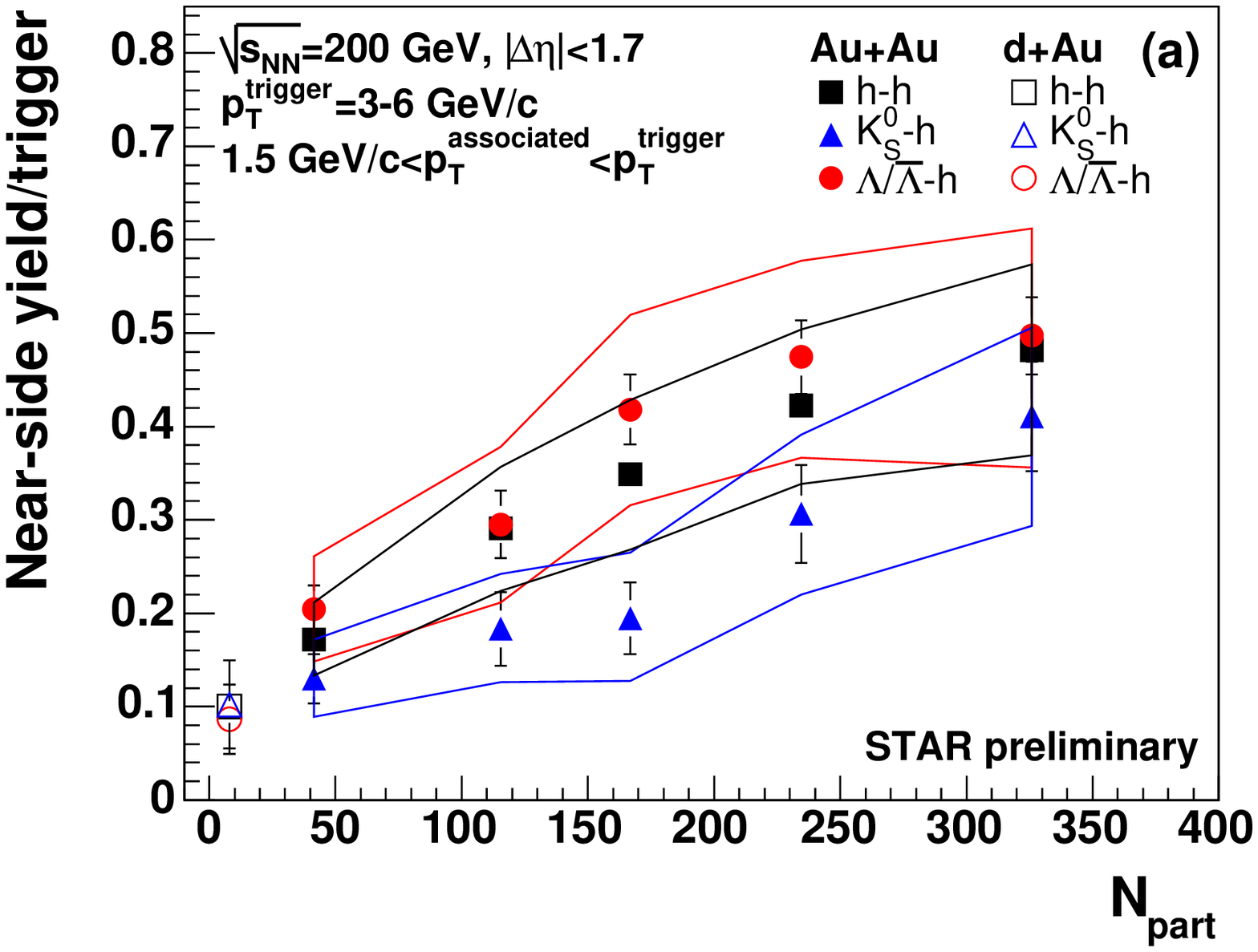,width=0.35\textwidth}%
\epsfig{file=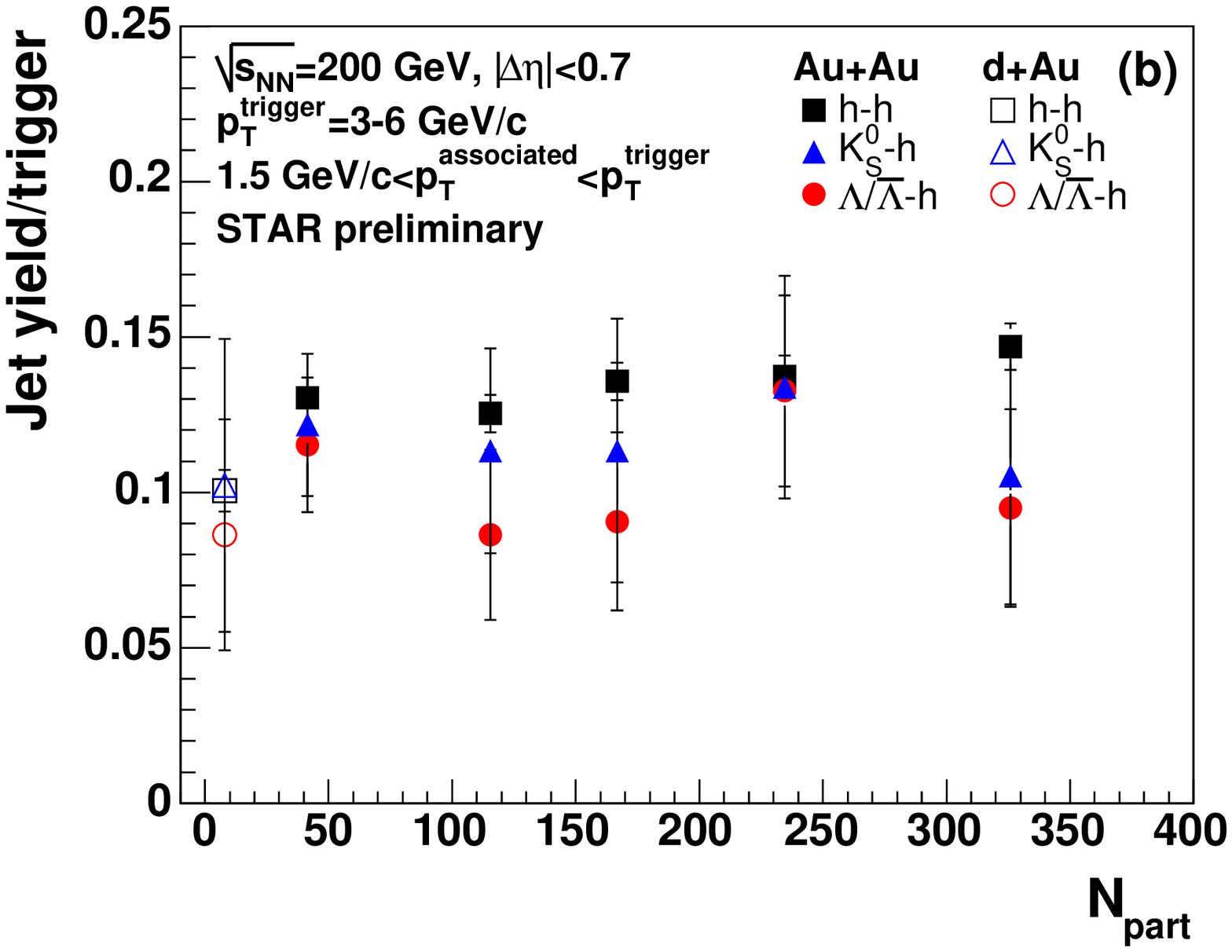,width=0.35\textwidth}
\caption{\label{fig:ridge_pid}Near-side correlated yield as a function of
  centrality for different trigger species. Left panel: Associated yield in
  full acceptance. Right panel: Jet yield from subtraction of the yield
  at large $\Delta\eta$ from the yield at small $\Delta\eta$.}
\end{figure}
A striking phenomenon in near-side jet-fragmentation was reported at this
conference: the increase in associated yield on the near side, which
is seen in Fig.\ \ref{fig:dihadr_star}, is found to have a large
contribution at larger $\Delta\eta$. The left Fig.\ \ref{fig:ridge_pid}
shows the distribution in \deta{} and \dphi{} of associated charged
particles with $\pTassoc > 2$ GeV with a trigger particle $3 < \pTtrig
< 4$ GeV \cite{putschke_qm06,bielcikova_qm06}. There is a clear elongation visible of the associated yield
on the near side. In fact, it seems that there are two components to
the near-side fragmentation: a central peak around
$(\Delta\eta,\Delta\phi)=(0,0)$ and a component that is elongated in
$\Delta\eta$, termed 'the ridge' by STAR \cite{putschke_qm06}. 

The size of the 'ridge'-effect is further illustrated in the middle
and right panels of Fig.\ \ref{fig:dihadr_star}. The middle panel
shows how the total yield on the near side increases with centrality
due to the ridge effect. The right panel shows the centrality
dependence of the near-side yield in the jet-like peak, which is
obtained by subtracting the yield measured at larger $\Delta\eta$ from
the yield at small $\Delta\eta$. The jet-like yield is independent of
centrality.

The analysis was performed for unidentified charged hadrons as well as
with leading identified $K^0$ and $\Lambda$. Both the jet and ridge
yields are similar for all trigger species, given the current rather
large statistical and systematic uncertainties. A similar effect is
seen in analyses of charged hadrons \cite{czhang_qm06} and identified
particles \cite{sickles_qm06} by PHENIX, but the effect is not as
pronounced due to the smaller $\eta$-acceptance.

\begin{figure}
\begin{minipage}{0.5\textwidth}
\caption{\label{fig:ridge_pt}Near-side associated hadron spectra for
  several \pTtrig{} in central Au+Au collisions. The associated yields
  are separated in a peak around $\Delta\eta = 0$ (jet-like
  contribution, dashed lines) and a $\Delta\eta$-independent contribution (the
  ridge, full lines).}
\end{minipage}\hfill
\begin{minipage}{0.45\textwidth}
\epsfig{file=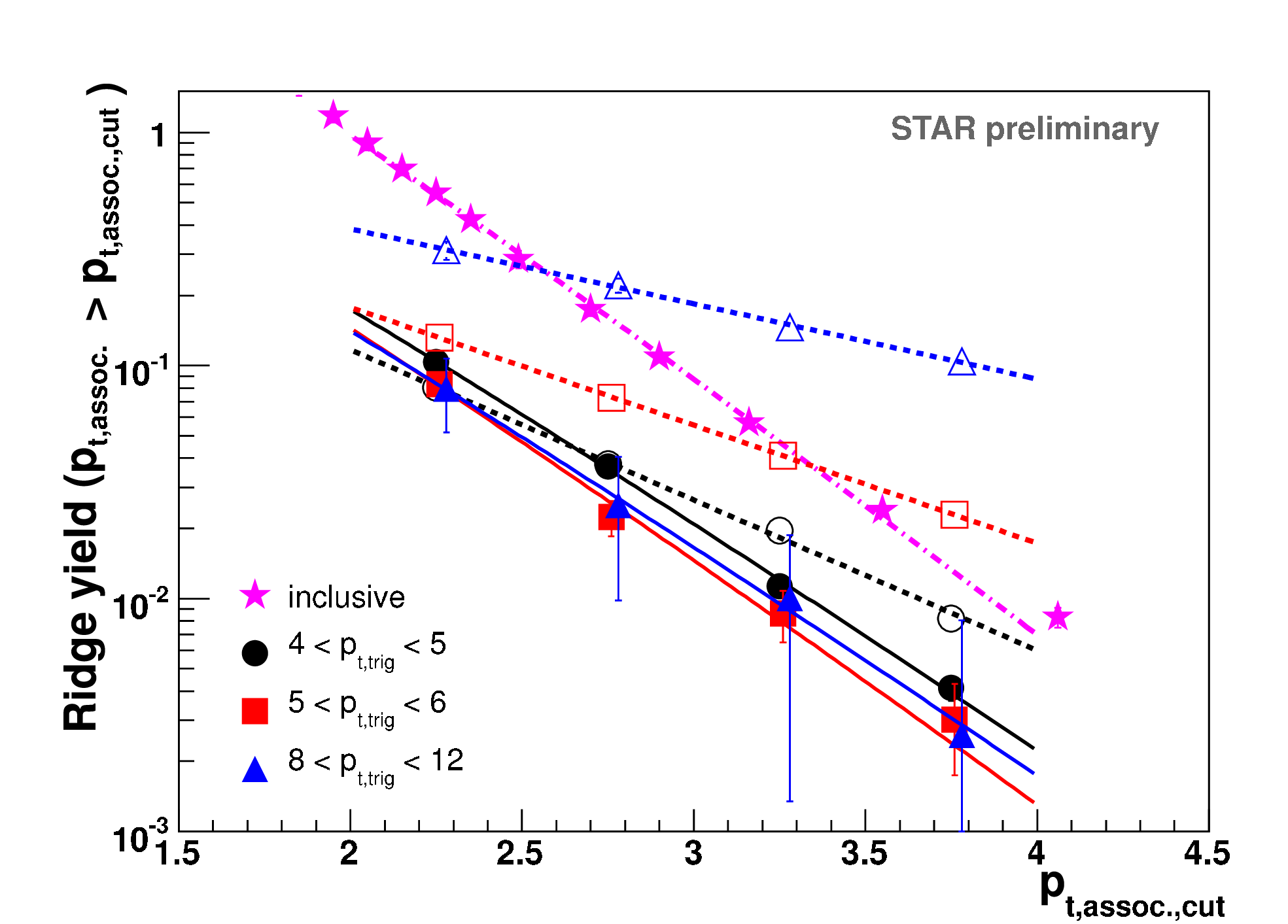,width=0.9\textwidth}
\end{minipage}
\end{figure}
Fig.\ \ref{fig:ridge_pt} shows how the jet and ridge yields evolve with
\pTassoc{} and \pTtrig. The jet component shows a clear increase with
\pTtrig, and a flattening of the \pTassoc{} spectrum as expected from
jet fragmentation. In fact, it turns out that the $z_T$-dependence of
the yields in the
jet-component are very similar to measurements in d+Au collisions
\cite{putschke_qm06}. The ridge-yield falls of more steeply with
\pTassoc, similar to the \pT-spectra from inclusive charged
particles. The ridge-yield is independent of \pTtrig, and remains
significant out to the highest measured \pTtrig.

It seems that the near-side jet-like peak is dominated by
fragmentation dynamics, while the ridge contains softer particles,
either from gluon radiation coupling to longitudinal flow or from the
flowing medium itself \cite{Armesto_flow,Voloshin_flow,Hwa_flow,Romatschke,Majumder}. 

\subsection{Away side shapes at low \pT: enhancement and broadening}
\begin{figure}
\epsfig{file=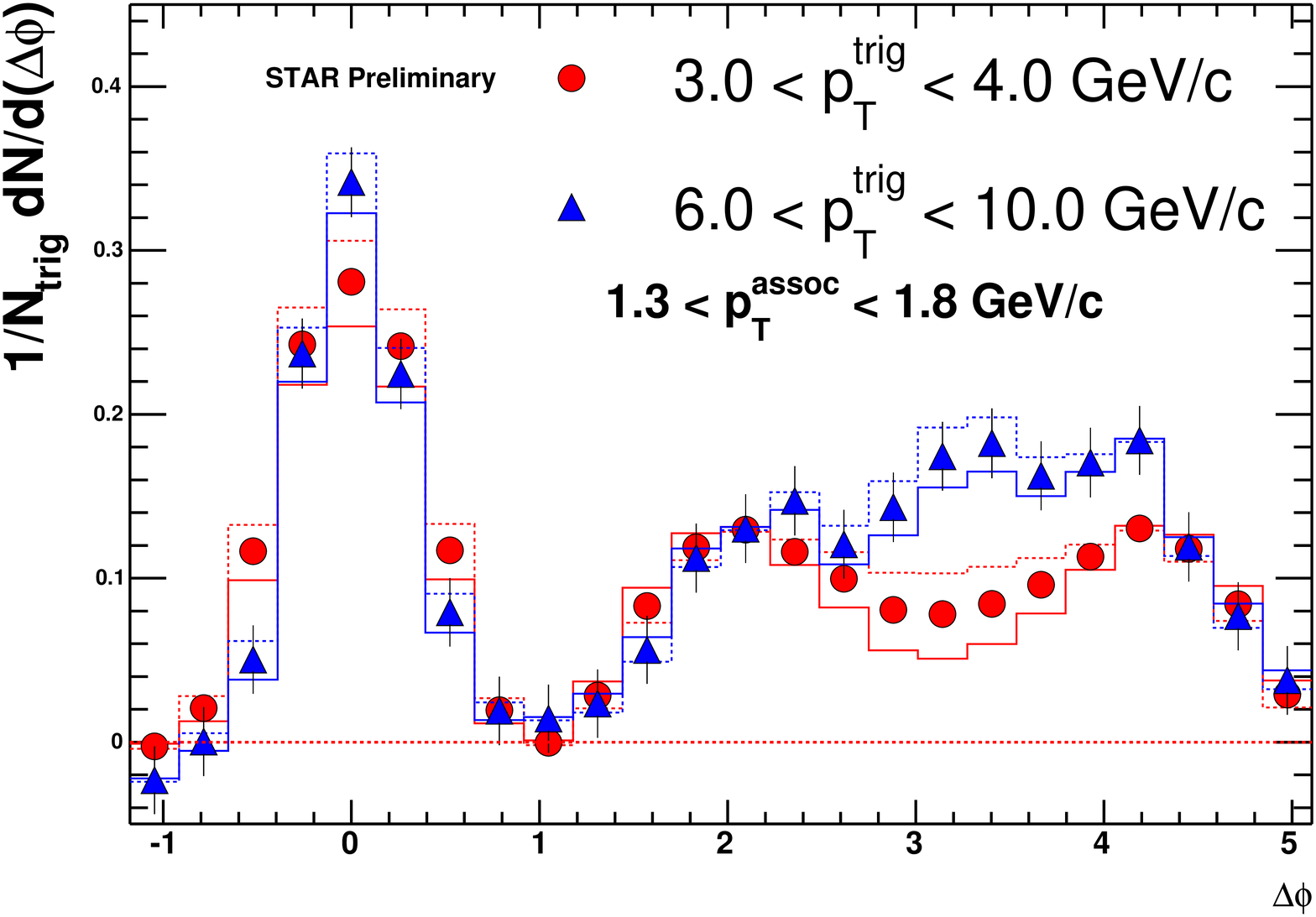,clip=,width=0.42\textwidth}\hfill
\epsfig{file=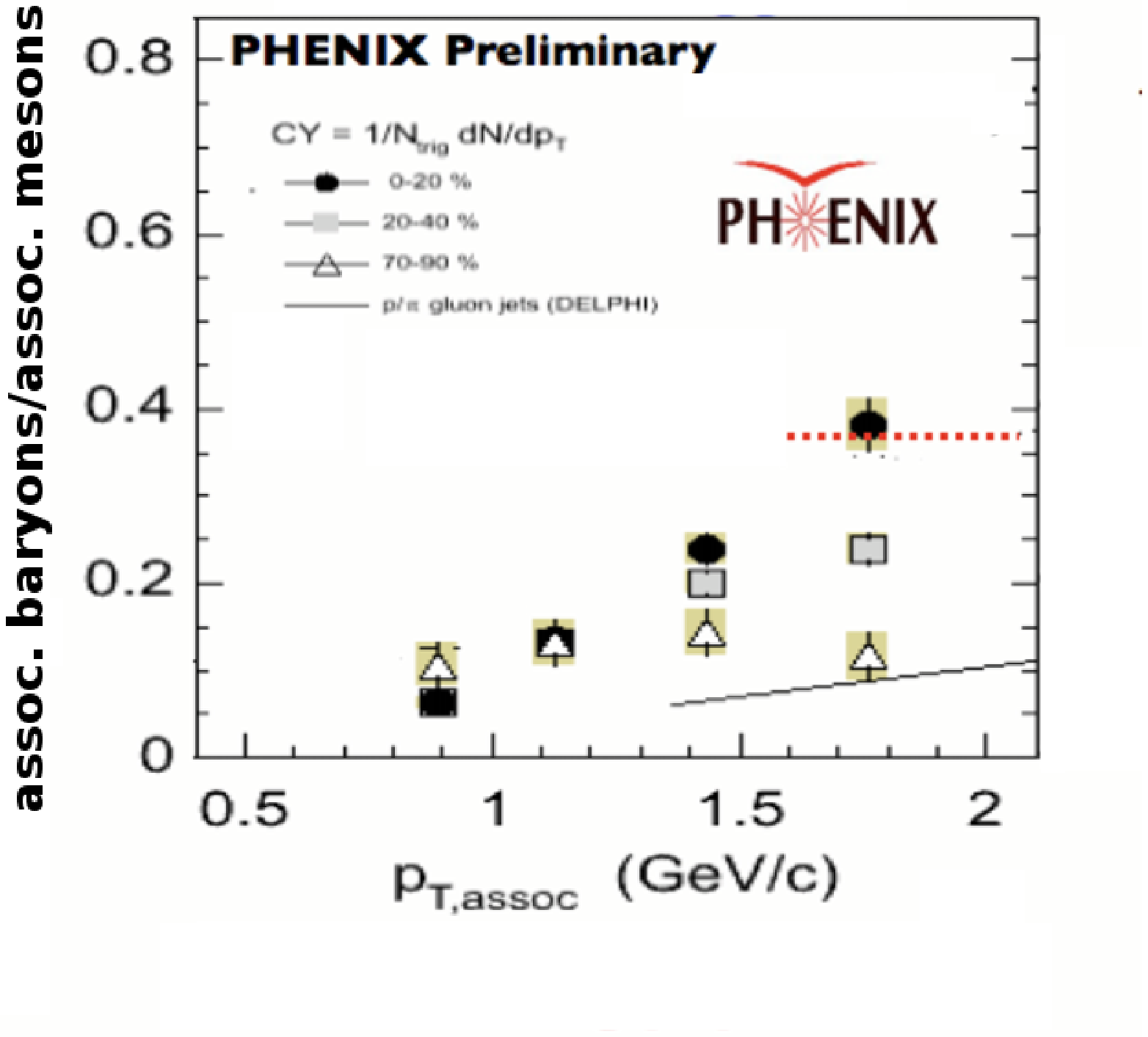,clip=,width=0.38\textwidth,bb=40 80
  500 470,clip=}
\caption{\label{fig:dihadr_int}Away-side properties at intermediate
  \pT. Left panel: Azimuthal distributions of hadrons with $1.3 < \pT
  < 1.8$ GeV associated with trigger hadrons in two different
  \pT-ranges, from 0-12\% central
  Au+Au collisions. Right panel: associated baryon/meson ratio on the away-side
  for charged trigger hadrons with $2.5 < \pT < 4.0$ GeV, for three
  different centralities in Au+Au collisions.}
\end{figure}
The away-side correlation shapes at low to intermediate \pT{} are also
found to be strongly modified in Au+Au collisions compared to p+p or
d+Au reference data. This is illustrated in Fig.\
\ref{fig:dihadr_int}, which shows azimuthal distributions of hadrons
with $1.3 < \pT < 1.8$ GeV associated with trigger hadrons in two
different \pT{} ranges from central 0-12\% Au+Au collisions
\cite{horner_qm06}. For lower \pTtrig{} ($3 < \pTtrig < 4$ GeV) a
depletion of the yield at $\Delta\phi=\pi$ is seen, leading to a
doubly-peaked structure. New data presented at this conference show
that for higher \pTtrig{} ($6 < \pTtrig < 10$ GeV, triangles in Fig.\
\ref{fig:dihadr_int}), there is no pronounced depletion. Clearly, the
shape and magnitude of the away-side associated yield evolve with \pTtrig{}
and \pTassoc{}, indicating the interplay of multiple processes which
dominate the dynamics in different kinematical regimes.

Detailed studies of the away-side structure at low and intermediate
\pT{} using three-particle correlations are being pursued by STAR and
PHENIX at RHIC and CERES at SPS
\cite{czhang_qm06,pruneau_qm06,kniege_qm06,fwang_qm06}. The aim of
these studies is to distinguish between scenarios involving conical emission 
\cite{solana_qm06} and large acoplanarity, but
other mechanisms that give rise to correlation structure, most
notably the effect of momentum conservation \cite{Borghini:2005kd},
should also be considered.

Explorations of jet structure with identified hadrons in the
intermediate \pT{} regime may also shed light on the baryon/meson
enhancement. The right panel of Fig.\ \ref{fig:dihadr_int} shows the
asssociated baryon/meson ratio on the away side for intermediate-\pT{}
trigger hadrons ($2.5 < \pTtrig < 4.0$ GeV) \cite{sickles_qm06}. For
central collisions, the baryon/meson ratio increases with \pTassoc{}
and reaches the value observed for inclusive charged particles
\cite{Adler:2003cb}. It would be interesting to see similar
measurements with trigger hadrons with $\pT \gtrsim 7$ GeV, where the
baryon/meson ratio reaches the value expected from jet fragmentation.

\section{Non-interacting probes: photons}

\begin{figure}
\begin{minipage}{0.5\textwidth}
  \caption{\label{fig:dir_gamma}Nuclear modification factor $R_{AA}$
  for direct photons in 0-10\% central Au+Au collisions, using new
  run-5 p+p result as a reference. The curves show various
  calculations of direct photon production, taking into account
  isospin effects and energy loss in the hot and dense medium.}
\end{minipage}
\hfill
\begin{minipage}{0.45\textwidth}
  \epsfig{file=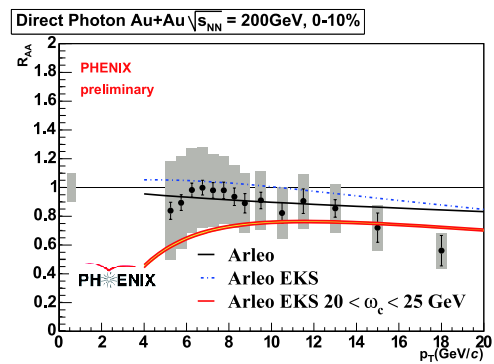,width=\textwidth}
\end{minipage}
\end{figure}

The increasing statistics being collected at RHIC for both heavy ion
and p+p events also leads to improved sensitivity for the rarest
probes, such as direct photons. Fig.\ \ref{fig:dir_gamma} shows the
nuclear modification factor for direct photons as measured by PHENIX,
now using new p+p results from RHIC run-5 \cite{isobe_qm06}. The
curves show a number of different expectations for $R_{AA}$
\cite{Arleo:2006xb}. A small suppression is expected from isospin
effects in the Au nucleus. A larger effect is expected due to the
suppression of fragmentation photons by energy loss of quark jets in
the medium (curves labeled $20 < \omega_c<25$ GeV). The data indicate
that there may be a small suppression at high \pT{}, but there is no
indication of the expected in-medium suppression at lower \pT, albeit
with large uncertainties.  It has also been suggested that
quark-photon conversions in the medium could lead to an enhancement of
direct photon production at intermediate \pT{}
\cite{Turbide:2005fk}. There is no evidence for such an enhancement in
the data.

\subsection{$\gamma$-hadron correlations}
\begin{figure}
  \epsfig{file=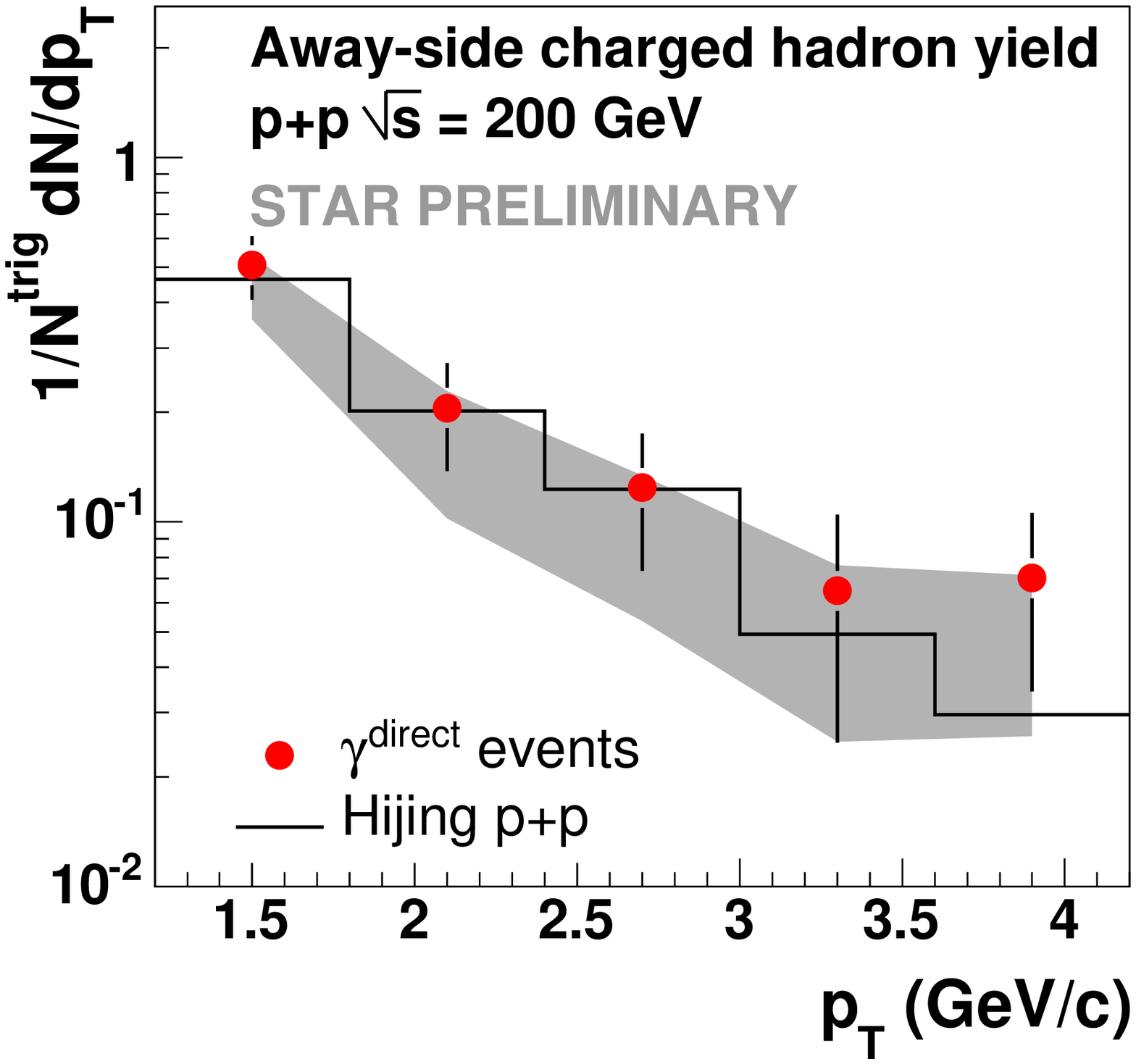,width=0.4\textwidth}\hfill
  \epsfig{file=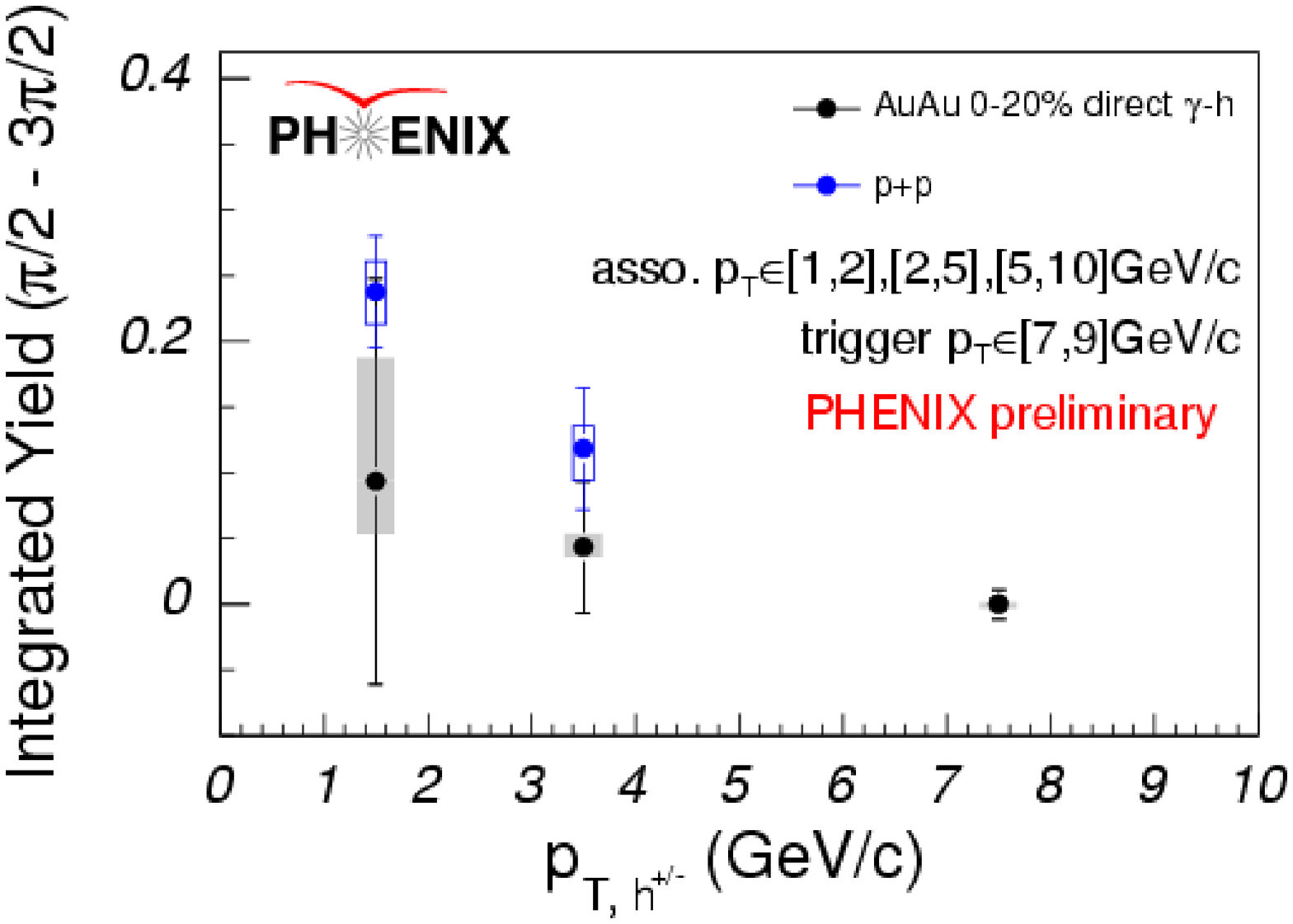,width=0.5\textwidth}
  \caption{\label{fig:gamma_jet}Away-side charged hadron spectra from
  events with a direct-photon trigger. Left panel: STAR results from
  p+p events with $8.5 < E_{T,trig} < 10.5$ GeV. Right panel: PHENIX
  results in p+p and Au+Au with $7 < E_{T,trig} < 9$ GeV.}
\end{figure}
High-\pT{} $\gamma$-jet production promises to be a sensitive probe of
parton energy loss, because it provides a sample of jets with
calibrated energy. This makes it possible to probe the probability
distribution of energy loss and thus the medium density profile, which
is very difficult with di-hadron measurements at RHIC
\cite{Renk:2006qg}. Fig.\ \ref{fig:gamma_jet} shows first results on
$\gamma$-jet measurements in p+p and Au+Au collisions from this
conference. The left panel shows the result from STAR in p+p
collisions \cite{subhasis_qm06}, which agrees with HIJING
expectations. The right panel shows results from PHENIX for p+p and
Au+Au collisions \cite{jin_qm06}. The Au+Au results are consistently
lower than the p+p results, as expected from parton energy loss. At
present, the statistical and systematic uncertainties are still
large. Both collaborations are aiming to improve the precision of the
Au+Au measurement with an upcoming large statistics Au+Au run.

\section{Conclusion and Outlook}
New results from high-statistics analyses at RHIC as presented at this
conference explore in-medium energy loss in detail. Measurements at
intermediate \pT{} show that there are large contributions to particle
production from phenomena that are not obviously related to vacuum jet
fragmentation, such as the near-side ridge and the strongly broadened
and enhanced away-side structure. Di-hadron correlation measurements
with identified particles are also being explored, and show for
example an increased baryon/meson ratio on the away side for
intermediate \pT{} trigger hadrons. So far, no clear connection
between the effects seen in correlation yields and the enhanced
baryon/meson ratio for inclusive particles has
been established, but this may happen in the near future, as more
detailed measurements with identified hadrons become available.

At higher \pT, particle production follows the
expectations of vacuum fragmentation more closely. With increasingly detailed
treatments of energy loss in model calculations, we are approaching a
quantitative description of the energy loss process and constraints on
the medium density profile. To constrain the theory further, comparisons
should be made with as much of the existing measurements as possible,
including nuclear modification factors, azimuthal asymmetries and
di-hadron fragmentation. In addition, centrality and system size dependence
measurements should be used to determine path length dependence and energy
density profiles.

$\gamma$-jet events provide a unique probe of parton energy loss
because the photon and the jet have balancing transverse momenta,
thus presenting a calibrated differential measure of energy loss. This
type of measurement is now actively being pursued at RHIC, with first
results from p+p collisions presented in this meeting. A next long Au+Au
run at RHIC will probably provide quantitative insight in parton
energy loss in $\gamma$-jet events. 

In the near future, first LHC data will
give a tremendous boost to the research with penetrating probes. The
kinematic range for measurements at LHC is well into the regime where
parton energies are larger than the typical energy loss, and this will
allow much more differential measurements. I am looking forward to
future results from LHC and from RHIC runs at higher luminosities which
will explore in-medium energy loss models and the medium density in
heavy ion collisions in quantitative detail.

\section*{References}
\bibliographystyle{epj}
\bibliography{mvanleeuwen_hipt_report}

\begin{thebibliography}{35}

\bibitem{Liu:2006ug}
H.~Liu, K.~Rajagopal, U.A. Wiedemann, Phys. Rev. Lett. \textbf{97}, 182301
  (2006), \texttt{hep-ph/0605178}\relax
\relax
\bibitem{star_LamK}
J.~Adams et~al. (STAR) (2006), \texttt{nucl-ex/0601042}\relax
\relax
\bibitem{Adler:2003cb}
S.S. Adler et~al. (PHENIX), Phys. Rev. \textbf{C69}, 034909 (2004),
  \texttt{nucl-ex/0307022}\relax
\relax
\bibitem{star_piprot}
B.I. Abelev et~al. (STAR), Phys. Rev. Lett. \textbf{97}, 152301 (2006),
  \texttt{nucl-ex/0606003}\relax
\relax
\bibitem{Adams:2005ph}
J.~Adams et~al. (STAR), Phys. Rev. Lett. \textbf{95}, 152301 (2005),
  \texttt{nucl-ex/0501016}\relax
\relax
\bibitem{Adler:2005ee}
S.S. Adler et~al. (PHENIX), Phys. Rev. Lett. \textbf{97}, 052301 (2006),
  \texttt{nucl-ex/0507004}\relax
\relax
\bibitem{Alt:2005cb}
C.~Alt et~al. (NA49), Nucl. Phys. \textbf{A774}, 473 (2006),
  \texttt{nucl-ex/0510054}\relax
\relax
\bibitem{Dainese:2005vk}
A.~Dainese (NA57), Nucl. Phys. \textbf{A774}, 51 (2006),
  \texttt{nucl-ex/0510001}\relax
\relax
\bibitem{reygers_qm06}
K.~Reygers et~al. (WA98) (2006), these proceedings,
  \texttt{nucl-ex/0701043}\relax
\relax
\bibitem{lajoie_qm06}
J.~Lajoie et~al. (PHENIX) (2006), these proceedings\relax
\relax
\bibitem{Dainese:2004te}
A.~Dainese, C.~Loizides, G.~Paic, Eur. Phys. J. \textbf{C38}, 461 (2005),
  \texttt{hep-ph/0406201}\relax
\relax
\bibitem{Wicks:2005gt}
S.~Wicks, W.~Horowitz, M.~Djordjevic, M.~Gyulassy (2005),
  \texttt{nucl-th/0512076}\relax
\relax
\bibitem{Salgado:2003gb}
C.A. Salgado, U.A. Wiedemann, Phys. Rev. \textbf{D68}, 014008 (2003),
  \texttt{hep-ph/0302184}\relax
\relax
\bibitem{Renk:2006qg}
T.~Renk, Phys. Rev. \textbf{C74}, 034906 (2006), \texttt{hep-ph/0607166}\relax
\relax
\bibitem{horner_qm06}
M.~Horner et~al. (STAR) (2006), these proceedings\relax
\relax
\bibitem{Renk:2006pk}
T.~Renk, K.J. Eskola (2006), \texttt{hep-ph/0610059}\relax
\relax
\bibitem{putschke_qm06}
J.~Putschke et~al. (STAR) (2006), these proceedings\relax
\relax
\bibitem{bielcikova_qm06}
J.~Bielcikova et~al. (STAR) (2006), these proceedings\relax
\relax
\bibitem{czhang_qm06}
C.~Zhang et~al. (PHENIX) (2006), these proceedings\relax
\relax
\bibitem{sickles_qm06}
A.~Sickles et~al. (PHENIX) (2006), these proceedings\relax
\relax
\bibitem{Armesto_flow}
N.~Armesto et~al., Phys. Rev. Lett. \textbf{93}, 242301 (2004),
  \texttt{hep-ph/0405301}\relax
\relax
\bibitem{Voloshin_flow}
S.A. Voloshin, Nucl. Phys. \textbf{A749}, 287 (2005),
  \texttt{nucl-th/0410024}\relax
\relax
\bibitem{Hwa_flow}
C.B. Chiu, R.C. Hwa, Phys. Rev. \textbf{C72}, 034903 (2005),
  \texttt{nucl-th/0505014}\relax
\relax
\bibitem{Romatschke}
P.~Romatschke, Phys. Rev. \textbf{C75}, 014901 (2007),
  \texttt{hep-ph/0607327}\relax
\relax
\bibitem{Majumder}
A.~Majumder, B.~Muller, S.A. Bass (2006), \texttt{hep-ph/0611135}\relax
\relax
\bibitem{pruneau_qm06}
C.~Pruneau et~al. (STAR) (2006), these proceedings\relax
\relax
\bibitem{kniege_qm06}
S.~Kniege et~al. (NA45/CERES) (2006), these proceedings\relax
\relax
\bibitem{fwang_qm06}
F.~Wang (2006), these proceedings\relax
\relax
\bibitem{solana_qm06}
J.~Casalderrey-Solana (2006), these proceedings\relax
\relax
\bibitem{Borghini:2005kd}
N.~Borghini, J.Y. Ollitrault, Phys. Lett. \textbf{B642}, 227 (2006),
  \texttt{nucl-th/0506045}\relax
\relax
\bibitem{isobe_qm06}
T.~Isobe et~al. (PHENIX) (2006), these proceedings\relax
\relax
\bibitem{Arleo:2006xb}
F.~Arleo, JHEP \textbf{09}, 015 (2006), \texttt{hep-ph/0601075}\relax
\relax
\bibitem{Turbide:2005fk}
S.~Turbide, C.~Gale, S.~Jeon, G.D. Moore, Phys. Rev. \textbf{C72}, 014906
  (2005), \texttt{hep-ph/0502248}\relax
\relax
\bibitem{subhasis_qm06}
S.~Chattopadhyay et~al. (STAR) (2006), these proceedings\relax
\relax
\bibitem{jin_qm06}
J.~Jin et~al. (PHENIX) (2006), these proceedings\relax
\relax
\end{thebibliography}

\end{document}